\documentclass[12pt]{article}
\usepackage{float}
\usepackage{authblk}
\usepackage{amssymb,amsmath}
\usepackage{graphicx}
\usepackage{caption}
\usepackage{subcaption}
\usepackage{cite}
\usepackage{color}
\usepackage{setspace}
\usepackage[colorlinks=true
,urlcolor=blue
,citecolor=blue
,linkcolor=blue
,pagecolor=blue
,linktocpage=true
,pdfproducer=medialab
]{hyperref}
\usepackage[a4paper,width=17cm,height=25cm]{geometry}
\makeatletter \renewcommand{\@dotsep}{10000} \makeatother
\title{Constraints on Two Higgs Doublet Model Parameters in the light of rare $B$-Decays}
\author[a] {Mureed Hussain%
	\thanks{\texttt mureed.hussain@sns.nust.edu.pk }}
\author[a] {Muhammad Usman%
	\thanks{\texttt muhammadusman@sns.nust.edu.pk}}
\author[a] {Muhammad Ali Paracha%
	\thanks{\texttt aliparacha@sns.nust.edu.pk}}
\author[b]{Muhammad Jamil Aslam%
	\thanks{\texttt muhammadjamil.aslam@gmail.com}}
\affil[a]{Department of Physics, School of Natural Sciences (SNS), National University of Sciences and Technology (NUST), Sector H-12 Islamabad, Pakistan.}
\affil[b]{Department of Physics, Quaid-i-Azam University, Islamabad, Pakistan.}
\begin{document}
\maketitle
\begin{abstract}
{We established the allowed parameters of two-Higgs doublet model(2HDM) from flavor physics observables, precisely from rare $B$ meson
	decays. In our analysis most formidable constraints on the 2HDM parameters arise from the branching ratio of rare radiative $B$ meson decay i.e.,
	$B\to X_{s}\gamma$. However, the constraints arising from the branching ratio of $B_{s}\to\mu^{+}\mu^{-}$ decay in $m_{H^{\pm}}-\tan\beta$ plane give 
	$m_{H^{\pm}} >$ 80 GeV for the value of $\tan\beta\sim 2$, that is in agreement with large electron-positron collider (LEP) data. 
	Furthermore, we also investigate the bounds on the $CP$-even $m_{H}$ and $CP$-odd $m_{A^{0}}$ Higgs boson not only from above mentioned physical observables, but also from the zero crossing of the forward-backward asymmetry of $B\to K^{\ast}\mu^{+}\mu^{-}$ decay. Therefore, these bounds on parameters of the 2HDM will
	provides a fertile ground to test the 2HDM at current and future $B$-physics experiments.}
	\end{abstract}

\newpage
\section{Introduction}
The discovery of Higgs boson by ATLAS and CMS experiments at the Large hadron collider (LHC)\cite{Chatrchyan:2012xdj, Aad:2012tfa} completes the only missing ingredient of Standard model (SM). In most of the cases, the results predicted by the SM are in good agreement with current experimental data, but still there are some unanswered questions that the SM cannot address, such as hierarchy problem, neutrino masses, dark matter, etc. In order to answer these questions, a number of new physics (NP) models have been proposed in literature. The NP signatures can be investigated through two possible approaches. In the first
approach, the direct observation, effects of the NP can be probed by smashing the particles at an adequate large energy and then explore the different particles produced as a result of this collision. The dedicated experiments for this purpose are the ATLAS and the CMS at the LHC. In the second approach, the NP effects can be explored via precision
studies especially in flavour physics and the devoted experiments for precision frontiers are the LHCb and Belle II at the super KEKB.	
The flavour physics processes are among the most suitable candles to explore the NP in the precision approach, especially the rare decays of $B$ and $K$ mesons. The rare $B$ meson decays are the ideal laboratory system to 
investigate NP as well as non-perturbtative aspects of QCD at low energy frontiers.  As mentioned earlier, in most of the cases, the predictions of the SM are in consensus
with current experimental data, but there are some anomalies at the level of $3\sigma$ observed in certain flavor physics observables. Just to mention, the LHCb results on the branching ratio of $B_{s} \to \phi \mu^{+}\mu^{-}$ in the two large-recoil bins deviates at the $3 \sigma$ level from its SM predictions \cite{Aaij:2013aln,Straub:2015ica}. Likewise, the LHCb analysis of the $3fb^{-1}$ of data on $B\to K^{*}\mu^{+}\mu^{-}$ confirms the anomaly (at  3$\sigma$ level) \cite{Aaij:2015oid} they have observed in the two large $K^{*}-$ recoil bins of angular observables $P^{\prime}_{5}$ \cite{DescotesGenon:2012zf, Descotes-Genon:2013vna,Altmannshofer:2013foa,Altmannshofer:2014rta,Altmannshofer:2017fio, Arnan:2017lxi} during their analysis of $1fb^{-1}$ data in 2013 \cite{Aaij:2013qta}. In addition, a measurement of the ratio of the branching fractions of the $B^{+} \to K^{+}\mu^{+}\mu^{-}$ and $B^{+} \to K^{+}e^{+}e^{-}$ shows a $2.6 \sigma$ deviations from the SM predictions \cite{Aaij:2014ora}. To resolve the issues that mimic in various observables in these decay modes, there exists a plenty of the NP models
such as model with extra dimensions \cite{ArkaniHamed:1998rs, Appelquist:2000nn}, little Higgs model \cite{ArkaniHamed:2002qy, Chang:2003vs}, family non-universal $Z^{\prime}$ models \cite{Langacker:2000ju, Barger:2009eq} and supersymmetric standard model \cite{Csaki:1996ks}. 
One of the most popular extensions of the SM is the two Higgs doublet model (2HDM) suggested by Lee \cite{Lee:1973iz} as a means to explain the matter-antimatter
asymmetry \cite{Gunion:1989we, Branco:2011iw}. A nice and comprehensive review of the 2HDM was presented in the ref. \cite{Branco:2011iw}. Just to be brief here, in the 2HDM, in addition to the SM Higgs doublet, an additional complex Higgs doublet was considered which then leads to two scalars $(h\;,H)$, one pseudo scalar $(A)$ and two charged $(H^{\pm})$ Higgs bosons. The vacuum expectation values (v.e.v) of the 2HDM are 
represented by $v_{1}$ and $v_{2}$ and the interactions of the fermions to the Higgs field, through which they acquire mass, depends on tangent of the ratio of the v.e.v ,
i.e., $\tan\beta=\frac{v_{2}}{v_{1}}$, and it serve as a free parameter in the 2HDM. In general, the 2HDM owns FCNC transition at tree level which can be avoided by imposing an ad-hoc discrete symmetry \cite{Glashow:1976nt}. Imposing an ad-hoc symmetry motivates to the two different possibilities in the 2HDM, namely the types I and II. In type I,  in order to retain the flavor conservation at tree level all the fermions couples with one of the Higgs doublet, whereas, in the type II scenario the 2HDM somehow harmonize with that of minimal supersymmetric model (MSSM) i.e., the up and down type quarks couples with two different Higgs doublets and so are the charged leptons. In addition to these two types, there are two other versions of the 2HDM in which the down type quarks and charged leptons acquire mass from different doublets and this refers to type III and type IV \cite{Barger:1989fj} and all these four types of 2HDM are summarize in Table \ref{Yukawa}. 	
From the experimental observation of branching ratios of $b\to s\gamma$ and the measurement of Higgs boson at the LHC \cite{BowserChao:1998yp, Dumont:2014wha}
one can get the indirect constraints on the masses of the 2HDM along with the constraint on $\tan\beta$.
This has been done in the past like Hou \textit{et. al.} \cite{Hou:1988gv} have discussed the charged Higgs boson effects on the loop induced $B-$ meson decays. Also in some recent studies (c.f. ref. \cite{Deschamps:2009rh}), a lower limit of 304 GeV on the charged Higgs mass in 2HDM Type-II is established by the branching ratio of $b \to s \gamma$. In refs. \cite{Cheng:2014ova,Cheng:2015yfu} constraints from $B_{s,d} \to \mu^{+}\mu^{-}$, $B \to \tau \nu$ and $B \to {X}_{s}\gamma$ are applied on charged Higgs boson and $\tan\beta$ in all four types of 2HDM. In \cite{Kling:2016opi} different search channels of exotic Higgs decays has been used to apply constraints on the 2HDM. In the present work we provide a comprehensive 
analysis of all types of 2HDM in light of rare $B$-decays. In particular we implement constraints on scalars, pseudo scalar, charged Higgs bosons masses and coupling 
parameters of the 2HDM from ALEPH collaboration on branching ratio of $b\to s\gamma$\cite{Barate:1998vz} and from LHCb collaboration results on the branching ratio of $B_{s}\to\mu^{+}\mu^{-}$ \cite{Aaij:2012nna} along with the measurement of the zero crossing of lepton forward-backward asymmetry $(A_{FB})$ of $B\to K^{\ast}\mu^{+}\mu^{-}$ \cite{Aaij:2013iag}.
\section{Theoretical Frame Work}
\subsection{Overview of Two Higgs Doublet Model}
In line with the SM of particle physics, the 2HDM has the only extension in Higgs sector where an extra Higgs doublet is introduced and it leaves all the other particle contents to be same. In general the scalar potential in the 2HDM has 11 independent parameters, but by imposing a particular symmetry will reduce the number of free parameters.
In most of the the 2HDM models a discrete $Z_{2}$ symmetry is imposed which takes first doublet $\Phi_{1}\to\Phi_{1}$ ($Z_{2}$ even) and the second doublet $\Phi_{2}\to-\Phi_{2}$ ($Z_2$ odd).
Implications of $Z_{2}$ symmetry left us with the 8 free parameters in $CP$-conserving potential, namely the four masses, the rotation angle in the CP-even sector, $\alpha$, the ratio of the vacuum expectation values, $\tan\beta = v_2/v_1$, and the soft breaking parameter $m^2_{12}$ and these along with their explicit form are explained in detail in ref.\cite{Arhrib:2013oia}. Just like the mass generation of fermions in the SM, in the 2HDM the fermions can
get mass because of Yukawa coupling $y_{ij}$ to Higgs doublet $\phi$. In the 2HDM, the Yukawa couplings to first scalar doublet $\Phi_{1}$ are fixed because the 
interactions have to be diagonal both in flavor space and in mass eigenstate basis. However, in case of second scalar doublet $\Phi_{2}$, the couplings are 
non-diagonal and can not be related to fermion masses. The mass eigenstates for fermions can be written as vectors in flavor space, therefore, the 2HDM Yukawa sector can
be expressed in terms of physical Higgs mass eigenstates as \cite{PhysRevD.72.035004}
	\begin{eqnarray}
	\mathcal{L}_{Yukawa}&=&-\dfrac{1}{\sqrt{2}}\bar{D}\{\kappa^D\sin(\beta-\alpha)+\rho^D\cos(\beta-\alpha)\}D~h- 
	\dfrac{1}{\sqrt{2}}\bar{D}\{\kappa^D\cos(\beta-\alpha)-\rho^D\sin(\beta-\alpha)\}D~H\nonumber\\ 
	&& -\dfrac{i}{\sqrt{2}}\bar{D}\gamma_5\rho^D D~A
	-\dfrac{1}{\sqrt{2}}\bar{U}
	\{\kappa^U\sin(\beta-\alpha)+\rho^U\cos(\beta-\alpha)\}U~h-\dfrac{1}{\sqrt{2}}\bar{U}
	\{\kappa^U\cos(\beta-\alpha)\nonumber\\
	&&-\rho^U\sin(\beta-\alpha)\}U~H- \dfrac{i}{\sqrt{2}}\bar{U}
	\gamma_5\rho^U U~A-\dfrac{1}{\sqrt{2}}\bar{L}\{\kappa^L\sin(\beta-\alpha)+
	\rho^L\cos(\beta-\alpha)\}L~h\nonumber\\
	&&-\dfrac{1}{\sqrt{2}}\bar{L}\{\kappa^L\cos(\beta-\alpha)
	-\rho^L\sin(\beta-\alpha)\}L~H-\dfrac{i}{\sqrt{2}}\bar{L}\gamma_5\rho^L L~A\nonumber\\
	&&-\left[\bar{U}\left(V_{CKM}\rho^DP_R-\rho^UV_{CKM}P_L\right)D~H^{+}+\bar{\nu}\rho^LP_RL~H^{+}+h.c\right].\label{1}
	\end{eqnarray}
In Eq. (\ref{1}), the $\kappa^{F}$ $(F = U\;, D\;, L)$'s are the $3\times3$ diagonal matrices with the definition $\kappa^{F}\equiv\sqrt{2}M^{F}/v$, where $M^{F}$' s are the corresponding fermion mass matrices. The detailed expressions of these matrices are given in reference \cite{Eriksson:2009ws}. The Lagrangian given in Eq. (\ref{1}) has a freedom to choose the arbitrary
value of $\rho^{F}$. However, the allowed size of the off-diagonal elements in $\rho^{F}$ have stringent constraint, because of non-zero elements instigate Higgs
mediated FCNC transition at tree level. Just for the sake of completeness, the connection between Yukawa coupling matrices $\rho^{F}$ and fermion mass matrices
$\kappa^{F}$ in four different types of 2HDM models \cite{Eriksson:2009ws} are summarized in Table \ref{Yukawa}. The purpose of the present study is to look for the constraints on all physical Higgs masses and coupling parameters in the light of the rare $B$-meson decays.
\begin{table}[h!]
\begin{center}
\begin{tabular}{ |c|c|c|c|c| }
\hline
& Type I & Type II & Type III & Type IV \\
\hline
$\rho_D$ & $\quad\kappa^D\cot(\beta)$ & $-\kappa^D\tan(\beta)$ & $-\kappa^D\tan(\beta)$ & $\quad\kappa^D\cot(\beta)$ \\ 
$\rho_U$ & $\quad\kappa^U\cot(\beta)$ & $\quad\kappa^U\cot(\beta)$ & $\quad\kappa^U\cot(\beta)$ & $\quad\kappa^U\cot(\beta)$ \\ 
$\rho_L$ & $\quad\kappa^L\cot(\beta)$ & $-\kappa^L\tan(\beta)$ & $\quad\kappa^L\cot(\beta)$ & $-\kappa^L\tan(\beta)$ \\ \hline
\end{tabular}
\end{center}
\caption{Relation between the fermion mass matrices and Yukawa coupling matrices in four 2HDM models.}
\label{Yukawa}
\end{table}
In the framework of the 2HDM type III Yukawa interaction, the Cheng-Sher-Yuan (CSY) parameterization \cite{Sher:1991km, PhysRevD:353484} for the couplings $\varepsilon_{ij}=\dfrac{m_i m_j}{v^2}\lambda_{ij}$ is useful and among these, the third family couplings $\lambda_{bb}$ and $\lambda_{tt}$ are constrained, along-with charged Higgs boson mass, from the branching ratios of $b \to s \gamma$ and $B^{+} \to l^{+} \nu$ \cite{Idarraga:2008zz}. They have shown that the most stringent constraints are coming from $b \to s \gamma$ and the pure leptonic decays can only exclude regions with small $\lambda_{bb}$ and small $m_{H^{\pm}}$. They have not used semileptonic rare B decays which we have applied in this article and also we use updated values of the branching ratios of radiative and pure leptonic $B$ decays. Here in the given notation of Yukawa couplings $\varepsilon_{ij}=\kappa^F\nu_{ij}+\rho^F\mu_{ij}$, where $\nu_{ij}$ and $\mu_{ij}$ relate the Higgs basis, $H$, and the generic basis, $\Phi$, as \cite{Haber:2006ue}
\begin{equation}
H_a=U_{ab}\Phi_b=
\begin{pmatrix}
\nu_{11}^{*} & \nu_{12}^{*} \\
\mu_{21}^{*} & \mu_{22}^{*}
\end{pmatrix}
\Phi_b=
\begin{pmatrix}
\nu_{11}^{*} & \nu_{12}^{*} \\
-\nu_{12} & \nu_{11}
\end{pmatrix}
\Phi_b \;.
\end{equation}

\subsection{Effective Hamiltonian}
\begin{figure}[H]
		\centering
		\begin{subfigure}{.35\textwidth}
			\centering
			\includegraphics[width=5cm ,height=2.5cm ]{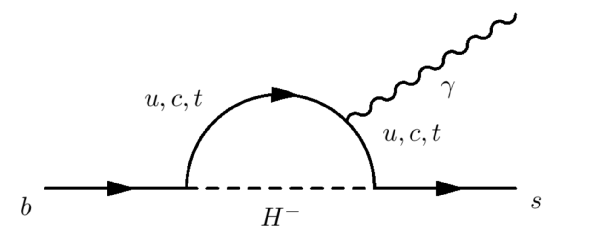}
			\caption{b $\rightarrow$ s $\gamma$}
			\label{fig:feyn1}
		\end{subfigure}%
		\begin{subfigure}{.3\textwidth}
			\centering
			\includegraphics[width=5.0cm , height=2.5cm]{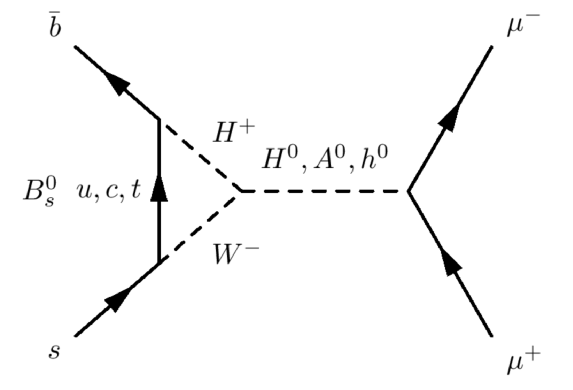}
			\caption{\textsc{$B_{s}$ $\rightarrow$ $\mu^{+}$ $\mu^{-}$}}
			\label{fig:feyn2}
		\end{subfigure}%
		\begin{subfigure}{.35\textwidth}
			\centering
			\includegraphics[width=4.5cm, height=2.5cm]{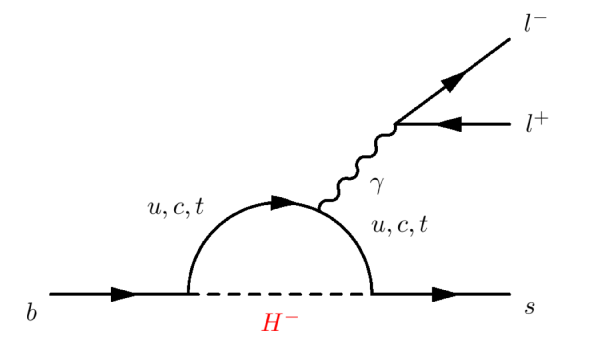}
			\caption{B $\rightarrow$ $K^{*} \mu^{+} \mu{-}$}
			\label{fig:feyn3}
		\end{subfigure}%
		\caption{Feynman diagrams of the three important rare decays involving $b -$quark.}
		\label{feyn3}
	\end{figure}

The phenomenology of the rare $B$-meson decays can be studied by using the effective Hamiltonian approach where one can separate the short distance physics (encoded in Wilson coefficients) from the long distance (concealed in transition form factors).  The effective Hamiltonian for rare radiative decay $b\to s\gamma$ and rare semileptonic
	decays $b \to s\ell^{+}\ell^{-}$ $(\ell = e\; ,  \mu\;, \tau )$ are given as follows \cite{Buras:1998raa}:
	\begin{equation}
	\begin{array}{rcl}
	\mathcal{H}_{eff}(b\to s\hspace{0.1cm}\gamma) = \dfrac{4\hspace{0.05cm}G_F}
	{\sqrt{2}}\displaystyle\sum_{p=u,c,t}^{}V_{ps}^{*}V_{pb}\sum\limits_{i=1}^{8}C_i(\mu)\hspace{0.05cm}O_i(\mu)~,\label{2}
	\end{array}
	\end{equation}
	\begin{equation}
	\begin{array}{rcl}
	\mathcal{H}_{eff}(b \to s\hspace{0.1cm}\ell^{+}\ell^{-}) = -\dfrac{4\hspace{0.05cm}G_F}
	{\sqrt{2}}V_{tb}V_{ts}^{*}\left(\displaystyle\sum\limits_{i=1}^{10}C_i(\mu)\hspace{0.05cm}O_i(\mu)+
	\sum\limits_{i=1}^{10}C_{Q_i}(\mu)\hspace{0.05cm}Q_i(\mu)\right)~.\label{3}
	\end{array}
	\end{equation}	
In Eqs. (\ref{2}) and (\ref{3}), $O_{i}(\mu)$ are the four quark local operators and $C_{i}(\mu)$ are the corresponding Wilson coefficients which are evaluated at 
energy scale $\mu$ which for the $B$-meson decays is the $b$-quark mass $(m_b)$.
The explicit form of the operators responsible for the decays of $B$-meson can be summarized as follows \cite{Chetyrkin:1997gb};
	\begin{eqnarray}\begin{array}{rcl}\label{operators}
	&& O_1=(\bar{s}\gamma_\mu T^a P_L c)(\bar{c} \gamma^\mu T^a P_L b),\text{\qquad}\hspace{1.3cm} 
	O_2=(\bar{s} \gamma_\mu P_L c)(\bar{c} \gamma^\mu P_L b)~, \nonumber \\
	&& O_3=(\bar{s}\gamma_\mu P_L b)\displaystyle\sum_{q}(\bar{q}\gamma^\mu q),\text{\qquad}\hspace{2cm}  O_4=
	(\bar{s}\gamma_\mu T^a P_L b)\displaystyle\sum_{q}(\bar{q}\gamma^\mu T^a q)~, \nonumber \\
	&& O_5=(\bar{s}\gamma_\mu\gamma_\nu\gamma_\rho P_L b)\displaystyle\sum_{q}(\bar{q}
	\gamma^\mu\gamma^\nu\gamma^\rho q),\text{\qquad}\hspace{0.40cm}  O_6=(\bar{s}\gamma_\mu\gamma_\nu\gamma_\rho T^a P_L b)
	\displaystyle\sum_{q}(\bar{q}\gamma^\mu\gamma^\nu\gamma^\rho T^a q)~,\nonumber \\
	&& O_7=\dfrac{e}{16\pi^2}[\bar{s}\sigma^{\mu\nu}(m_s P_L+m_b P_R)b]F_{\mu\nu},\text{\qquad}\hspace{-0.2cm}  O_8=\dfrac{g}
	{16\pi^2}[\bar{s}\sigma^{\mu\nu}(m_s P_L+m_b P_R)T^a b]G^a_{\mu\nu}~, \nonumber \\
	&& O_9=\dfrac{e^2}{(4\pi)^2}(\bar{s}\gamma^\mu b_L)(\bar{\ell}\gamma_\mu \ell),\text{\qquad}\hspace{1.9cm}
	O_{10}=\dfrac{e^2}{(4\pi)^2}(\bar{s}\gamma^\mu b_L)(\bar{\ell}\gamma_\mu\gamma_5 \ell) ~,\nonumber 
	\end{array}
	\end{eqnarray}
	\begin{eqnarray}
	\begin{array}{rcl}
	&& Q_1=\dfrac{e^2}{16\pi^2}(\bar{s}_L^\alpha b_R^\alpha)(\bar{\tau}\tau),\text{\qquad}\hspace{2.45cm}  Q_2=
	\dfrac{e^2}{16\pi^2}(\bar{s}_L^\alpha b_R^\alpha)(\bar{\tau}\gamma_5\tau)~, \\
	&& Q_3=\dfrac{g^2}{16\pi^2}(\bar{s}_L^\alpha b_R^\alpha)\displaystyle\sum_q(\bar{q}_L^\beta q_R^\beta),
	\text{\qquad}\hspace{1.4cm}  Q_4=\dfrac{g^2}{16\pi^2}(\bar{s}_L^\alpha b_R^\alpha)\displaystyle\sum_q(\bar{q}_R^\beta q_L^\beta)~, \\
	&& Q_5=\dfrac{g^2}{16\pi^2}(\bar{s}_L^\alpha b_R^\beta)\displaystyle\sum_q(\bar{q}_L^\beta q_R^\alpha),\text{\qquad}\hspace{1.4cm}
	Q_6=\dfrac{g^2}{16\pi^2}(\bar{s}_L^\alpha b_R^\beta)\displaystyle\sum_q(\bar{q}_R^\beta q_L^\alpha)~, \\
	&& Q_7=\dfrac{g^2}{16\pi^2}(\bar{s}_L^\alpha\sigma^{\mu\nu} b_R^\alpha)\displaystyle\sum_q
	(\bar{q}_L^\beta \sigma_{\mu\nu} q_R^\beta),\text{\qquad}\hspace{0.25cm}  Q_8=\dfrac{g^2}{16\pi^2}
	(\bar{s}_L^\alpha\sigma^{\mu\nu} b_R^\alpha)\displaystyle\sum_q(\bar{q}_R^\beta \sigma_{\mu\nu} q_L^\beta)~, \\
	&& Q_9=\dfrac{g^2}{16\pi^2}(\bar{s}_L^\alpha\sigma^{\mu\nu} b_R^\beta)\displaystyle\sum_q(\bar{q}_L^\beta 
	\sigma_{\mu\nu} q_R^\alpha),\text{\qquad}\hspace{0.25cm}  Q_{10}=\dfrac{g^2}{16\pi^2}(\bar{s}_L^\alpha\sigma^{\mu\nu} 
	b_R^\beta)\displaystyle\sum_q(\bar{q}_R^\beta \sigma_{\mu\nu} q_L^\alpha) ~.\label{4}
	\end{array}
	\end{eqnarray}
In Eq. (\ref{4}) the operators $O_{i}(i=1,. . .,10)$ are in the SM basis, whereas the new operators $Q_{i}\; (i=1,. . .,10)$ correspond to the contributions from
neutral Higgs bosons (NHBs) exchange diagrams and they are depicted in Fig. \ref{feyn3}. The explicit form of the Wilson coefficients for NHBs are given below \cite{Dai:1996vg}:
	\begin{eqnarray*}
		C_{Q_{1}}(m_{W})&=& \frac{m_{b}m_{\ell}}{m_{h^{0}}^2}\tan^{2}\beta \frac{1}{\sin^{2}\theta_{W}}\frac{x}{4}\times
		\bigg\{(\sin^{2}\alpha+h\cos^{2}\alpha)f_{1}(x,y)\\
		&&+\Big[\frac{m_{h^{0}}^2}{m^{2}_{W}}+(\sin^{2}\alpha+h\cos^{2}\alpha)(1-z)\Big]f_{2}(x,y)
		-\frac{\sin^{2}2\alpha}{4}\frac{(m_{h^{0}}^2-m_{H^{0}}^2)^{2}}{m_{h^{0}}^2m_{H^{0}}^2}\bigg\}\notag\\
		C_{Q_{2}}(m_{W})&=& -\frac{m_{b}m_{\ell}}{m_{A^{0}}^2}\tan^{2}\beta \frac{1}{\sin^{2}\theta_{W}}\frac{x}{4}\times\bigg\{f_{1}(x,y)+
		\Big(1-\frac{m_{H^{\pm}}^2-m_{A^{0}}^2}{m^{2}_{W}}\Big)f_{2}(x,y)\bigg\}
		\end{eqnarray*}
		\begin{eqnarray*}
		C_{Q_{3}}(m_{W})&=&\frac{m_{b}e^{2}}{m_{\ell}g^{2}}\Big(C_{Q_{1}}(m_{W})+C_{Q_{2}}(m_{W})\Big)\\
		C_{Q_{4}}(m_{W})&=&\frac{m_{b}e^{2}}{m_{\ell}g^{2}}\Big(C_{Q_{1}}(m_{W})-C_{Q_{2}}(m_{W})\Big)\\
		&&C_{Q_{i}}(m_{W})=0,  i=5,. . .,10.
	\end{eqnarray*}
     where 
	  $x=m_t^2 / m_W^2$,
	  $y=m_t^2 / {m_H^{\pm}}^2$,      	
	$h= m_h^2 / m_H^2$ and
	$z=x / y.$
	\section{Scanning Technique For 2HDM Parameters and Experimental Constraints from Rare $B$-Decays}	
	In order to study the constraints on the parameters of the 2HDM we use a two Higgs doublet model 
	calculator (2HDMC) which is a C++ code and is based on object oriented programming \cite{Eriksson:2010zzb}. In
	2HDMC package one has a choice to tune the Higgs potential parameters and also in Yukawa sector it
	gives a freedom to enumerate the couplings which leads to the FCNC. In this package a standard choices of
	Yukawa couplings were used and here Type-I to IV represent different Yukawa couplings 
	as demonstrated in Table \ref{Yukawa}. The 2HDMC also examines the theoretical properties of the 2HDM
	, such as the unitarity of the $S$-matrix and positivity of the potential. As described earlier, in the present study we perform the random
	scan on the 2HDM physical basis parameters such as $\{m_{h}\;,m_{H}\;,m_{A}\;,m_{H^{\pm}}\;,m^{2}_{12}\;,
	\tan\beta\;, \sin(\beta-\alpha)\;,\lambda_{6}\;,\lambda_{7};\}$, in the following range: 
	\begin{eqnarray}
	124.0 \leq   &m_h& \leq 126.0 \text(GeV) \nonumber \\
	0 \leq    &m_H& \leq 1000   \text(GeV) \nonumber \\
	0 \leq    &m_A& \leq 1000   \text(GeV) \label{s1} \\  
	0 \leq   &m_{H^{\pm}}& \leq 1000   \text(GeV)  \nonumber \\
	-5000 \leq   &m^{2}_{12}& \leq 5000   \text(GeV)  \nonumber \\
	0 \leq   &\tan\beta&   \leq 10  \nonumber \\
	- 1 \leq   &\sin(\beta - \alpha)&  \leq  1\nonumber
	\end{eqnarray}
where $m_h$ is the SM like Higgs boson, while $m_{H}$, $m_{A}$ and $m_{H^{\pm}}$ are the CP-even,
 CP-odd and the charged Higgs bosons, respectively. 
 $m^{2}_{12}$ is a free parameter in the Yukawa Lagrangian of the 2HDM as defined in \cite{Eriksson:2009ws} and
 $\tan\beta$ is the ratio of the vacuum expectation values of the two Higgs doublets.	
Making use of the $Z_{2}$ symmetry on the Yukawa Lagrangian we set $\lambda_{6}=\lambda_{7}=0$.
 As mentioned above our goal is to investigate the 2HDM parameters in light of rare $B$ decays,
 for this purpose we embed 2HDMC on SuperIso v3.4 \cite{Mahmoudi:2008tp} to study the flavor physics
 observables such as, branching 
ratios of $b\to s\gamma, B_{s}\to\mu^{+}\mu^{-}$ and zero crossing of the forward-backward asymmetry 
of $B\to K^{\ast}\mu^{+}\mu^{-}$. We then use values of these observables to constraint 
 the 2HDM parameter space. 	 
 Following are experimental values of the $BR(B\to X_s\gamma)$, $BR(B_{s}\to\mu^{+}\mu^{-})$ and
 the zero crossing $q^{2}_{0}$ of the forward-backward asymmetry $A_{FB}$ of 
$B \rightarrow K^{*} \mu^{+} \mu^{-} $,  are used to constraint the 2HDM parameters.
\begin{eqnarray}
 BR(B \rightarrow X_{s} \gamma) &=& (3.36 \pm 0.23) \times 10^{-4} \nonumber \\
 BR(B_s \rightarrow \mu^{+} \mu^{-}) &=& 3.0^{+1.0}_{-0.9} \times 10^{-9}   \\
  q^{2}_{0} &=& 4.9 \pm 0.9 \nonumber
 \end{eqnarray}
 \section{Results and Analysis}
\begin{figure}[H]
 	\begin{subfigure}{.5\textwidth}
 		\centering
 		\caption{Type I $m_{H^{\pm}}$ vs tan$\beta$}   
 		\includegraphics[width=1.0\linewidth]{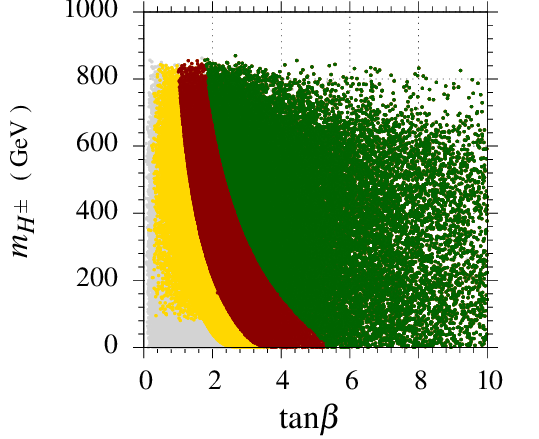}
 		\label{fig5:sfig1}
 	\end{subfigure}%
 	\begin{subfigure}{.5\textwidth}
 		\centering
 		\caption{Type II $m_{H^{\pm}}$ vs tan$\beta$}   
 		\includegraphics[width=1.0\linewidth]{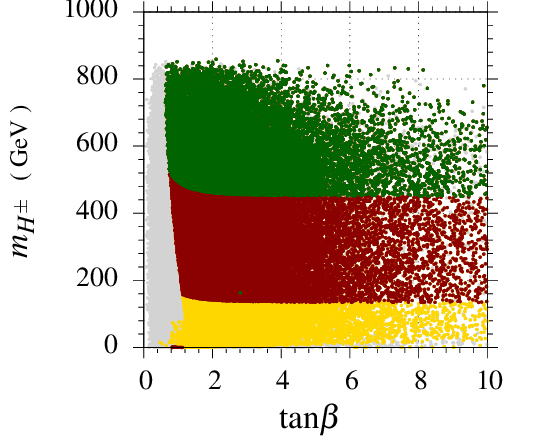}
 		\label{fig5:sfig2}
 	\end{subfigure}%
 	\\
 	\begin{subfigure}{.5\textwidth}
 		\centering
 		\caption{Type III $m_{H^{\pm}}$ vs tan$\beta$}   
 		\includegraphics[width=1.0\linewidth]{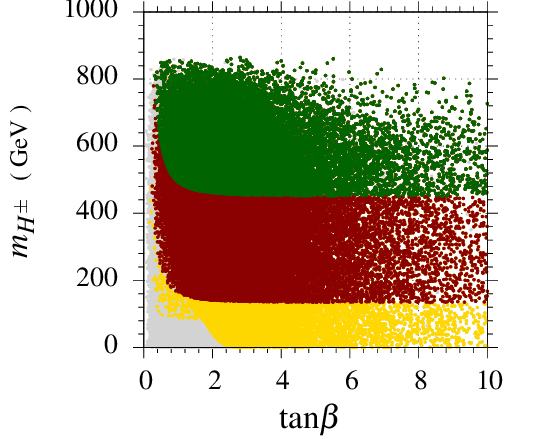}
 		\label{fig5:sfig3}
 	\end{subfigure}%
 	\begin{subfigure}{.5\textwidth}
 		\centering
 		\caption{Type IV $m_{H^{\pm}}$ vs tan$\beta$}   
 		\includegraphics[width=1.0\linewidth]{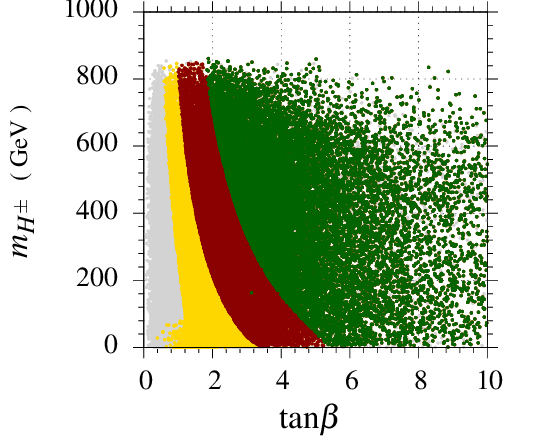}
 		\label{fig5:sfig4}
 	\end{subfigure}%
 	\caption{Effect of constraints on tan$\beta$ vs $m_{H^{\pm}}$ plane that stem from the branching ratio of $B\rightarrow \mu^{+} \mu^{-}$, the zero crossing ($q^{2}_{0}$) of forward-backward asymmetry of $B \rightarrow K^{*} \mu^{+} \mu^{-}$ and branching ratio of $b \rightarrow s \gamma$ is shown in yellow, red and green colors respectively.}
 	\label{fig:combined}
 \end{figure}

 In this section, we present the scan over the 2HDM parameter space given in Eq. (\ref{s1}).
 In all figures, the Gray region is consistent with the unitarity of the $S$-matrix and positivity 
 of the potential of the 2HDM. Yellow region is subset of the Gray region by satisfying the constraint
 from $BR(B_s \rightarrow \mu^{+} \mu^{-})$, whereas the Red region subset the yellow region and it
 satisfy the constraints from the zero position of  $A_{FB}$ in $B \rightarrow K^{*} \mu^{+} \mu^{-} $. Green region is
  subset of red region and satisfies the constraint from $BR(B \rightarrow X_s \gamma)$.  
  Figure \ref{fig:combined} shows the effect of above mentioned decays on the 
  $m_{H^{\pm}}-\tan\beta$ plane of the 2HDM.
  The most stringent constraints in all types 
  of 2HDM arise from $b \rightarrow s \gamma$ decay. The second important constraint is from $A_{FB}$ of $B \rightarrow K^{*} \mu^{+} \mu^{-} $, whereas the effects of  $B_{s}$ $\rightarrow$ $\mu^{+} \mu^{-}$ on $m_{H^{\pm}}- \tan\beta$ in all 2HDM planes are minimal. In general, the upper limit on $m_{ H ^ {\pm}}$ in 2HDM is 840 GeV (c.f. Fig. \ref{fig:combined}) . 
  Fig. \ref{fig5:sfig1} represents Type-I of 2HDM. In this figure one can see that for low values of
  $\tan\beta$ ($\tan\beta \sim 2$),  the value of  $m_{ H ^ {\pm}}$ should be greater than $80$ GeV which is consistent with the LEP data \cite {Agashe:2014kda} and also with the value of $m_{ H ^ {\pm}}$ that is constrained from 
  $ BR (B_{s} \rightarrow \mu^{+} \mu^{-})$. From the zero crossing of $A_{FB}$ of 
  $B \rightarrow K^{*} \mu^{+} \mu^{-} $  the allowed value of $\tan\beta$ $>$ $2.5$ but this value of $\tan\beta$ reduces to $1$ when we increase $m_{H^{\pm}}$ up to $800$ GeV. 
  The constraints from $BR(B \to X_s \gamma)$ implies that $\tan\beta \leq 4.5$ is not allowed for $m_{H^{\pm}}$
  $\sim$ $80$ GeV but this lower bound also decreases when we increase $m_{H^{\pm}}$ and for $m_{H^{\pm}} \sim$ 800 GeV the value of $\tan\beta \sim 2.0$.    \\
  In Fig. \ref{fig5:sfig2} we present Type-II of the 2HDM in which $\tan\beta \leq 1$ is not 
  allowed for all above mentioned constraints. We can see from this figure 
  that $m_{H ^{\pm}}<125$ GeV is not allowed from the constraint of zero crossing of $A_{FB}$.
  However, the things get more interesting from the constraints of branching ratio of the decay
  $B \rightarrow  X_s \gamma$ as in this case, the lower limit of $m_{H^{\pm}} \sim 460$ GeV and
  so the allowed band for $m_{H ^{\pm}}$ narrows 
  (460 GeV $\leq$ $m_{H ^{\pm}}$ $\leq$ 840 GeV). This is in accordance with the bounds given on $m_{H ^{\pm}}$ by doing NNLO calculations for $B \to X_s \gamma$ decay by Misiak \emph{et. al} \cite{Misiak:2015xwa}.\\
 In Fig. \ref{fig5:sfig3} we discuss the Type-III of the 2HDM in which the effect of constraints on
 $m_{H ^{\pm}}$ are almost similar to that of Type-II. The only difference is for 
 $m_{H ^{\pm}}< 85$ GeV, which anyhow is not allowed by LEP. Similarly, most of the limits on these two parameters in Type-IV of the
 2HDM are in accordance with that of the Type-I. The possible differences are only for $\tan\beta<1$ and $m_{H ^{\pm}}< 85$ GeV which in any case is out of the possible allowed region.  \\
 The similarity in above results is due to the fact that for pure leptonic decay the Yukawa coupling
 $\rho_{L}$ is same for Type-I and Type-III so the trends 
 of constraints from branching ratio of $B_{s} \to \mu^{+} \mu^{-}$ are same in these two types. Likewise, the Yukawa coupling  $\rho_{L}$ is same in Type-II and Type-IV. 
 In contrast to this, the Yukawa coupling $\rho_{D}$ is same in Type-I and Type-IV and likewise for Type-II and Type-III versions of 2HDM. Therefore, the constraints from zero crossing of $A_{FB}$
 and from the branching ratio  of $B \to X_s \gamma $ are same for Type-I and Type-IV while they 
 are similar for Type-II and Type-III.
 \begin{figure}[H]
 \begin{subfigure}{.5\textwidth}
   \centering
\caption{Type I $m_{H}$ vs $\tan\beta$}
   \includegraphics[width=1.0\linewidth]{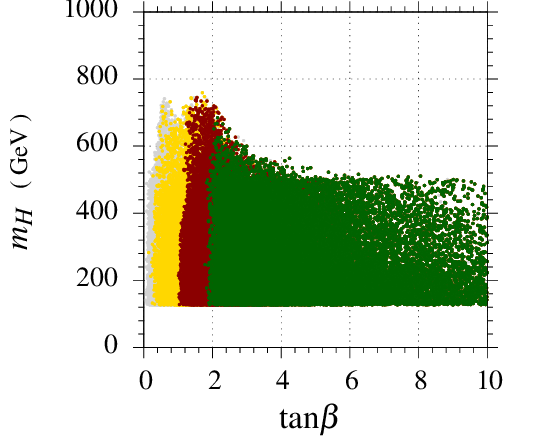} 	
   \label{fig6:sfig1}
 \end{subfigure}%
 \begin{subfigure}{.5\textwidth}
   \centering
   \caption{Type II $m_{H}$ vs $\tan\beta$}
   \includegraphics[width=1.0\linewidth]{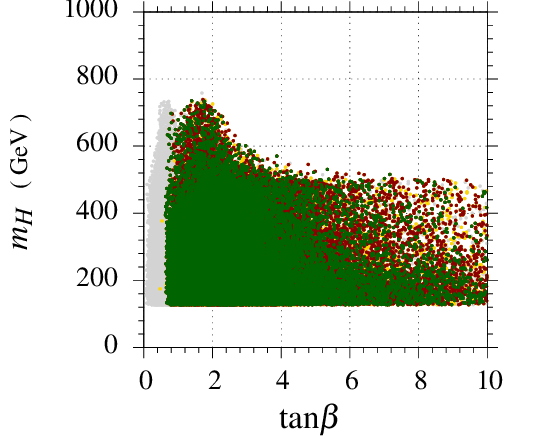}
   \label{fig6:sfig2}
 \end{subfigure}%
 \\
 \begin{subfigure}{.5\textwidth}
   \centering
   \caption{Type III $m_{H}$ vs $\tan\beta$}
   \includegraphics[width=1.0\linewidth]{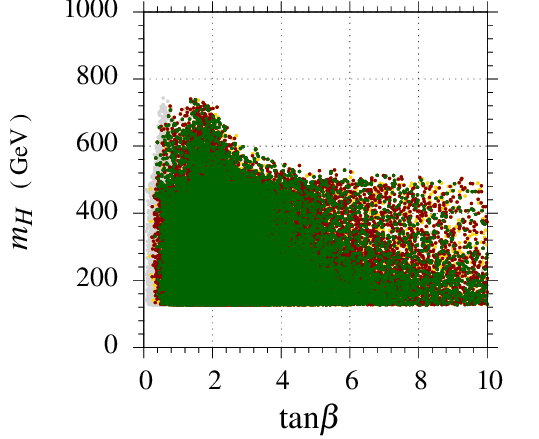}
   \label{fig6:sfig3}
 \end{subfigure}%
 \begin{subfigure}{.5\textwidth}
   \centering
   \caption{Type IV $m_{H}$ vs $\tan\beta$}
   \includegraphics[width=1.0\linewidth]{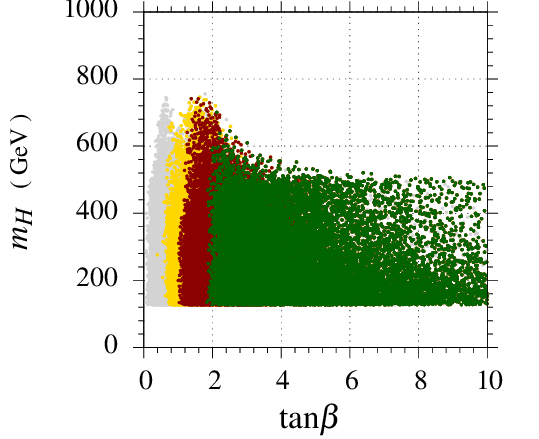}
   \label{fig6:sfig4}
 \end{subfigure}%
 \caption{Effect of constraints on $m_{H}$ vs $\tan\beta$ plane. Color coding is same as in 
 figure \ref{fig:combined}.}
 \label{fig:6}
 \end{figure}
 In Fig.\ref{fig:6} we display our results in $m_H -\tan\beta$ planes. From these figures, one can see that while increasing the value of $\tan\beta$ the upper limit of $m_{H}$ decreases. To be precise, for the value of $\tan\beta \sim 6$,  $m_{H}$ can not be higher than $500$ GeV. 
 We also infer the effects of constraints from the above mentioned rare $B-$~meson decays which are similar in Type-I and Type-IV of 2HDM whereas these constraints have no effect on Type-II and Type-III.  Also from Fig. \ref{fig6:sfig1} and Fig. \ref{fig6:sfig4},
 the value of $\tan\beta \leq 1$ is not allowed from constraints of zero crossing of $A_{FB}$ and similar to this $\tan\beta \leq 2$ is forbidden from constraints of $BR(B \to X_s \gamma)$ decay.
  In Fig.\ref{fig:7}, we show the behavior of a pseudo-scalar Higgs boson $(m_{A^0})$ with  $\tan\beta$. As long as the mass of pseudo-scalar Higgs is concerned, one can notice that almost all the mass range of $m_{A^{\circ}}$ is allowed in Type-I and Type-III versions of the 2HDM. But at the same time, we can see that $\tan\beta<2$ is not allowed in Type-I (c.f. Fig. \ref{fig7:sfig1}). From Fig.~\ref{fig7:sfig2}, the lower bound on $m_{A^{\circ}}$ can be predicted to be
  $60$ GeV and $120$ GeV from constraints of zero crossing of $A_{FB}$ and  $BR (B \to X_s \gamma)$, respectively.
  
 \begin{figure}[H]
 \begin{subfigure}{.5\textwidth}
   \centering
 	\caption{Type I $m_{A^{o}}$ vs $\tan\beta$}
   \includegraphics[width=1.0\linewidth]{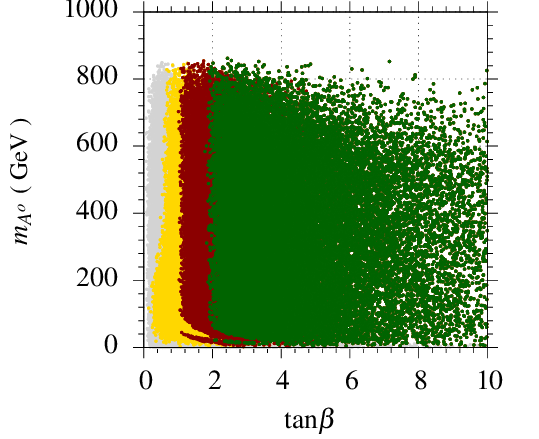}
   \label{fig7:sfig1}
 \end{subfigure}%
 \begin{subfigure}{.5\textwidth}
   \centering
   \caption{Type II $m_{A^{o}}$ vs $\tan\beta$}
   \includegraphics[width=1.0\linewidth]{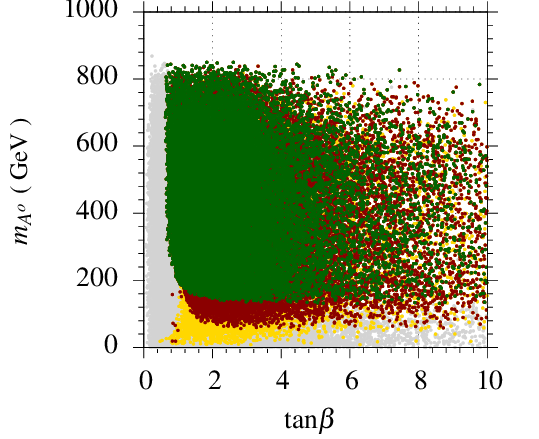}
   \label{fig7:sfig2}
 \end{subfigure}%
 \\
 \begin{subfigure}{.5\textwidth}
   \centering
   \caption{Type III $m_{A^{o}}$ vs $\tan\beta$}
   \includegraphics[width=1.0\linewidth]{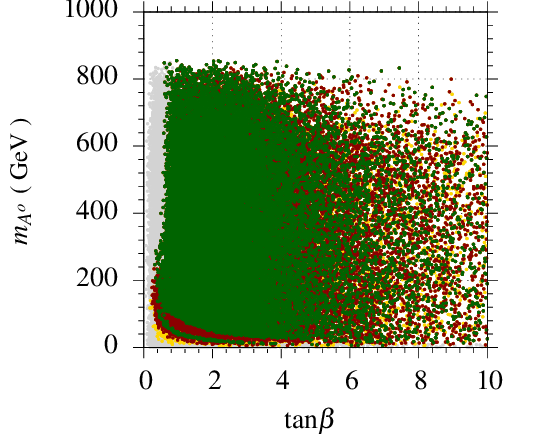}
   \label{fig7:sfig3}
 \end{subfigure}%
 \begin{subfigure}{.5\textwidth}
   \centering
   \caption{Type IV $m_{A^{o}}$ vs $\tan\beta$}
   \includegraphics[width=1.0\linewidth]{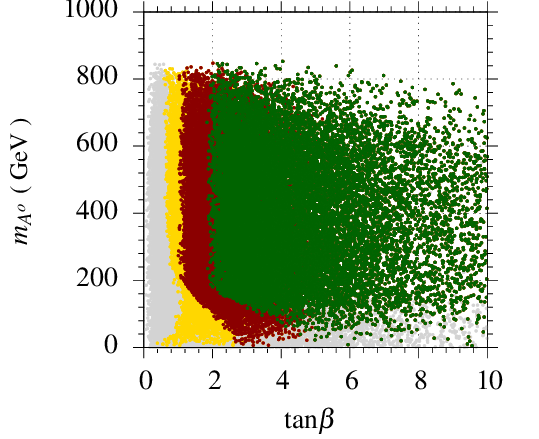}
   \label{fig7:sfig4}
 \end{subfigure}%
 \caption{Effect of constraints on $m_{A^{o}}$ vs $\tan\beta$ plane. Color coding is same as in
 figure \ref{fig:combined}.}
 \label{fig:7}
 \end{figure}
\begin{figure}[H]
 \begin{subfigure}{.5\textwidth}
   \centering
 	\caption{Type I $\lambda_{tt}$ vs $m_{H^{\pm}}$}
   \includegraphics[width=1.0\linewidth]{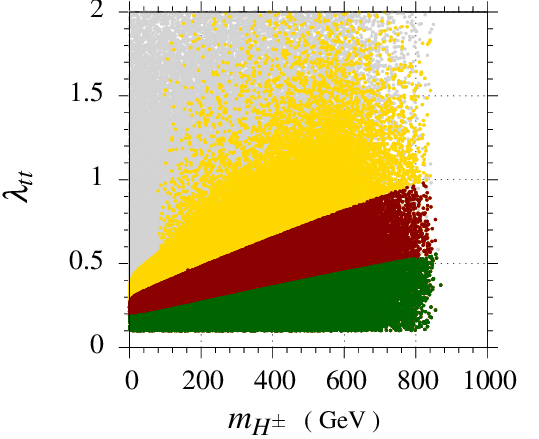}
   \label{fig10:sfig1}
 \end{subfigure}%
 \begin{subfigure}{.5\textwidth}
   \centering
   \caption{Type II $\lambda_{tt}$ vs $m_{H^{\pm}}$}
   \includegraphics[width=1.0\linewidth]{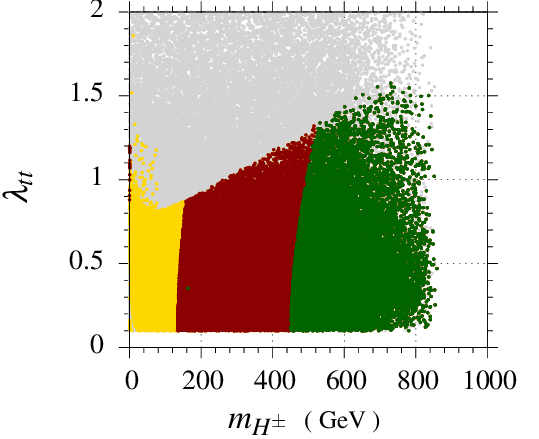}
   \label{fig10:sfig2}
 \end{subfigure}%
 \\
 \begin{subfigure}{.5\textwidth}
   \centering
   \caption{Type III $\lambda_{tt}$ vs $m_{H^{\pm}}$}
   \includegraphics[width=1.0\linewidth]{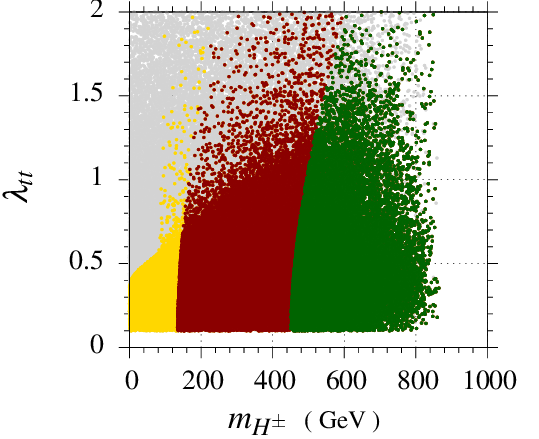}
   \label{fig10:sfig3}
 \end{subfigure}%
 \begin{subfigure}{.5\textwidth}
   \centering
   \caption{Type IV $\lambda_{tt}$ vs $m_{H^{\pm}}$}
   \includegraphics[width=1.0\linewidth]{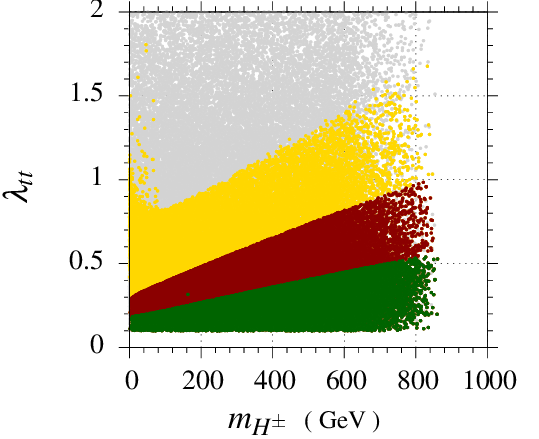}
   \label{fig10:sfig4}
 \end{subfigure}%
 \caption{Effect of constraints on $\lambda_{tt}$ vs $m_{H^{\pm}}$ plane. Color coding is same as in
 figure \ref{fig:combined}.}
 \label{fig:10}
 \end{figure}

  Now let us discuss the behavior of coupling $\lambda_{tt}$ with the mass of charged Higgs boson $(m_{H^{\pm}})$ in all four types of the 2HDM.
  As we have already mentioned that the most stringent constraints on the masses of the 2HDM parameters 
  stem from the $BR(B \to X_s \gamma)$ and the least constraint is coming from the $BR(B_s \to \mu^{+} \mu^{-})$.
  This trend keeps on for the couplings $\lambda_{tt}$ and $|\lambda_{bb}|$. 
  In Fig. \ref{fig10:sfig1}, it can be observed that the constraints on $\lambda_{tt}$ from branching ratio of 
  $B_s$ $\rightarrow$ $\mu^{+}$ $\mu^{-}$ allows almost all values of $\lambda_{tt}$. However, things
  become different when we constraint these two parameters from the other two observables. For 
  example, one can see that the allowed range of $\lambda_{tt}$, from the constraint coming from zero 
  crossing of $A_{FB}$, linearly increases from 0.3 to 1 as we increase
  $m_{H^{\pm}}$. 
  From the constraints of $BR(B \to X_s \gamma)$, the allowed range for coupling $\lambda_{tt}$ gets 
  more restricted and the upper limit is 0.55 when $m_{H^{\pm}}\approx 800$ GeV.  
  $80$ GeV the range of allowed values of $\lambda_{tt}$ increases linearly with the maximum value of 1.55 at 
  $m_{{H^{\pm}}} \approx 800$ GeV.
  In Fig. \ref{fig10:sfig3} we can see that the constraints on $m_{{H^{\pm}}}$ are almost similar 
  to those in Type-II. Likewise, the trend of  $\lambda_{tt}$ and $m_{{H^{\pm}}}$ is similar to Type-I in the Type-IV version of the 2HDM (c.f. Fig. \ref{fig10:sfig4} ).
  
\begin{figure}[H]
 \begin{subfigure}{.5\textwidth}
   \centering
         \caption{Type I $\lambda_{tt}$ vs $m_{H}$}
   \includegraphics[width=1.0\linewidth]{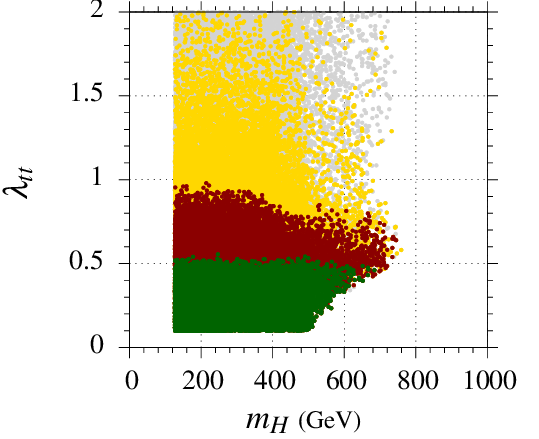}
   \label{fig11:sfig1}
 \end{subfigure}%
 \begin{subfigure}{.5\textwidth}
   \centering
   \caption{Type II $\lambda_{tt}$ vs $m_{H}$}
   \includegraphics[width=1.0\linewidth]{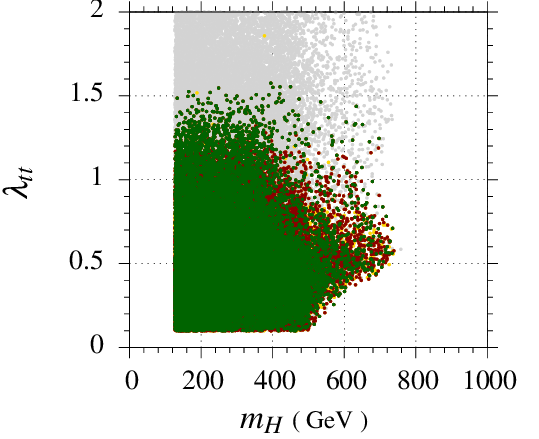}
   \label{fig11:sfig2}
 \end{subfigure}%
\\
 \begin{subfigure}{.5\textwidth}
   \centering
   \caption{Type III $\lambda_{tt}$ vs $m_{H}$}
   \includegraphics[width=1.0\linewidth]{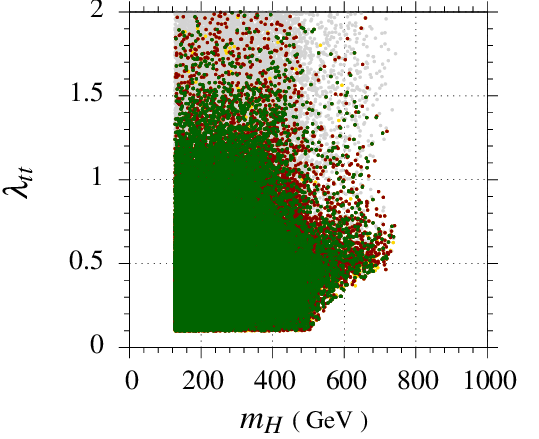}
   \label{fig11:sfig3}
 \end{subfigure}%
 \begin{subfigure}{.5\textwidth}
   \centering
   \caption{Type IV $\lambda_{tt}$ vs $m_{H}$ }
   \includegraphics[width=1.0\linewidth]{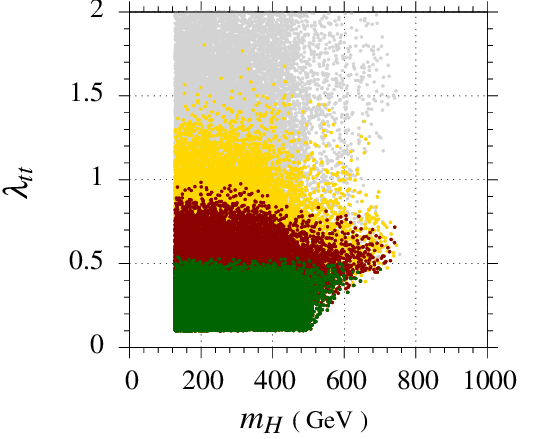}
   \label{fig11:sfig4}
 \end{subfigure}%
 \caption{Effect of constraints on $\lambda_{tt}$ vs $m_{H}$ plane. Color coding is same as in
 Fig. \ref{fig:combined}}
 \label{fig:11}
 \end{figure}

\begin{figure}[H]
 \begin{subfigure}{.5\textwidth}
   \centering
         \caption{Type I $\lambda_{tt}$ vs $m_{A^{o}}$}
   \includegraphics[width=1.0\linewidth]{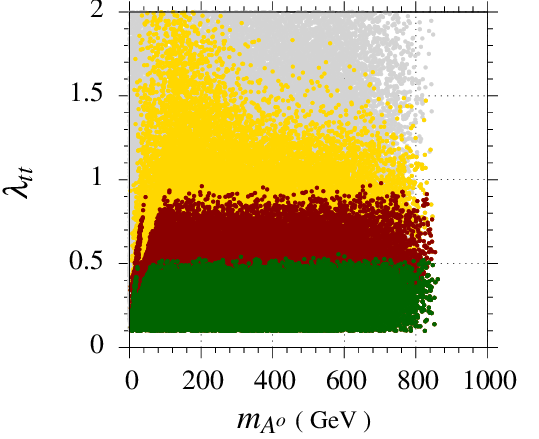}
   \label{fig12:sfig1}
 \end{subfigure}%
 \begin{subfigure}{.5\textwidth}
   \centering
   \caption{Type II $\lambda_{tt}$ vs $m_{A^{o}}$}
   \includegraphics[width=1.0\linewidth]{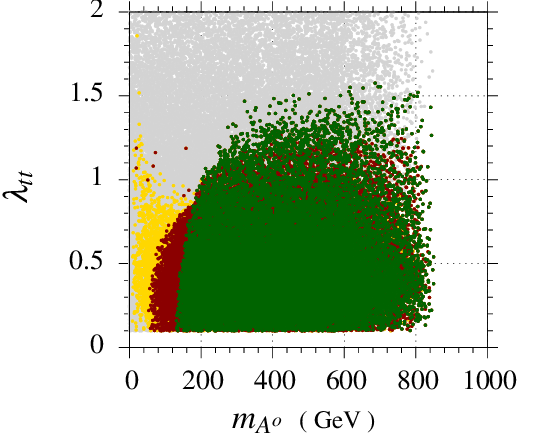}
   \label{fig12:sfig2}
 \end{subfigure}%
\\ 
 \begin{subfigure}{.5\textwidth}
   \centering
   \caption{Type III $\lambda_{tt}$ vs $m_{A^{o}}$}
   \includegraphics[width=1.0\linewidth]{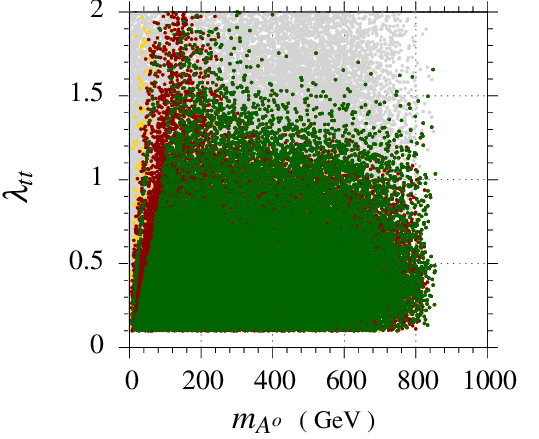}
   \label{fig12:sfig3}
 \end{subfigure}%
\begin{subfigure}{.5\textwidth}
   \centering
   \caption{Type IV $\lambda_{tt}$ vs $m_{A^{o}}$ }
   \includegraphics[width=1.0\linewidth]{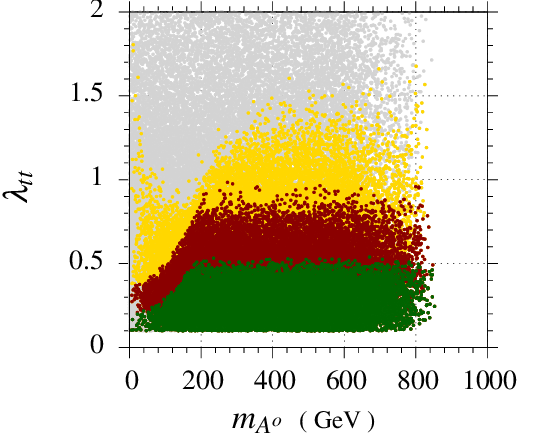}
   \label{fig12:sfig4}
 \end{subfigure}%
 \caption{Effect of constraints on $\lambda_{tt}$ vs $m_{A^{o}}$ plane. Color coding is same as in
 Fig. \ref{fig:combined}}
 \label{fig:12}
 \end{figure}

  Another interesting thing is to look for the allowed region of  $\lambda_{tt}$ when it is plotted 
  against the mass of $CP$ even neutral Higgs boson $(m_{H})$ in all the four Yukawa types of the 2HDM. In case 
  of the Type-I and Type-IV of the 2HDM as depicted in Fig.~\ref{fig11:sfig1} and \ref{fig11:sfig4}, respectively for 
  $m_{H} \leq 500$ GeV the value of $\lambda_{tt}$ can not be greater
  then 1 when constrained from zero crossing of $A_{FB}$. Also we can see that $\lambda_{tt}>0.5$ is 
  not allowed from constraints arising due to
  $BR(B \to X_s \gamma)$. On the other hand, in case of the Type-II and Type-III of the 2HDM, as displayed in Fig. \ref{fig11:sfig2} and \ref{fig11:sfig3}, respectively ,
  there is no constraint on these parameters form the input $B$-meson decays. Also for higher values of $m_{H}$
  i.e., for  $m_{H} \simeq 700$ GeV, $\lambda_{tt}<0.5$ is not allowed in all the four types of the 2HDM.
In Fig. \ref{fig:12}, we display variation of $\lambda_{tt}$ with mass of pseudo-scalar Higgs boson $(m_{A^{o}})$ . In case of the Type-I and
Type-IV, the constraints on $\lambda_{tt}$ from $q^2_{0}$ and $b \to s \gamma$ are similar. For example, taking into account the constraints from
$q^2_{0}$, for $m_{A^{\circ}}<200$ GeV there is a linear increase in the $\lambda_{tt}$ up-to 
$\lambda_{tt} \simeq 1$ and for $m_{A^{\circ}}>200 GeV$, $\lambda_{tt}$ can attain the value up-to 1.
Likewise, from $b \to s \gamma$, for $m_{A ^{\circ}}\leq160$ GeV, there is again a linear(almost) increase in 
the allowed range of $\lambda_{tt}$ that can go to $\lambda_{tt}\simeq 0.5$ in this mass range. However, for the rest of the mass 
range $\lambda_{tt}$ can have any value less than $0.5$. Now from $B_{s}\to \mu^{+} \mu^{-}$ by looking at the trend of $\lambda_{tt}$, it can be noticed that this decay does not give any particular effects in case of the Type-I of the 2HDM. 
For Type-IV the allowed range 
of $\lambda_{tt}$ increases as we increase $m_{A^{\circ}}$ but the upper limit for $\lambda_{tt}$ 
for this type is $\lambda_{tt}\simeq 1.75$. 
However, for Type-II of the 2HDM, in Fig. \ref{fig12:sfig2}, $\lambda_{tt}$ can not be greater than $1.6$, whereas for 
the Type-III as plotted
in Fig. \ref{fig12:sfig3}, there is no bound on $\lambda_{tt}$ for 
$m_{A^{\circ}}\leq 400$ GeV and for $m_{A^{\circ}}> 400$ GeV, $\lambda_{tt}$ should be less than 
$1.75$.

\begin{figure}[H]
 \begin{subfigure}{.5\textwidth}
   \centering
 	\caption{Type I $|\lambda_{bb}|$ vs $m_{H^{\pm}}$}
   \includegraphics[width=1.0\linewidth]{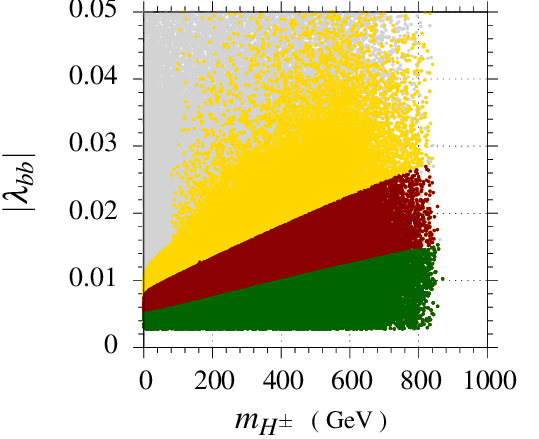}
   \label{fig13:sfig1}
 \end{subfigure}%
 \begin{subfigure}{.5\textwidth}
   \centering
  \caption{Type II $|\lambda_{bb}|$ vs $m_{H^{\pm}}$}
   \includegraphics[width=1.0\linewidth]{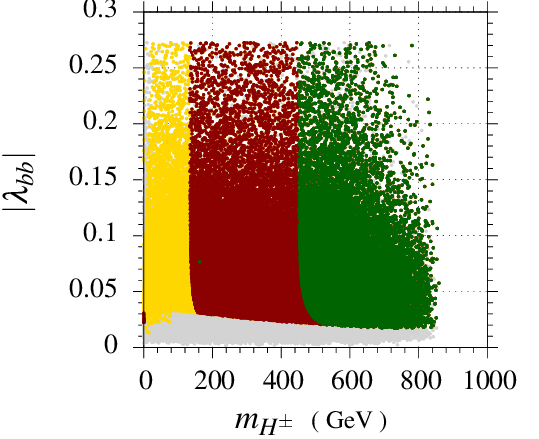}
   \label{fig13:sfig2}
 \end{subfigure}%
 \\
 \begin{subfigure}{.5\textwidth}
   \centering
   \caption{Type III $|\lambda_{bb}|$ vs $m_{H^{\pm}}$}
   \includegraphics[width=1.0\linewidth]{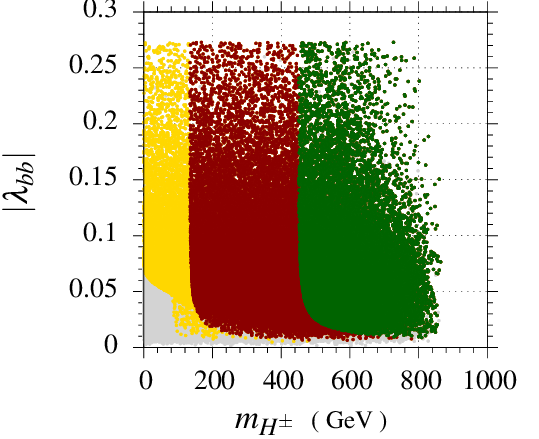}
   \label{fig13:sfig3}
 \end{subfigure}%
 \begin{subfigure}{.5\textwidth}
   \centering
   \caption{Type IV $|\lambda_{bb}|$ vs $m_{H^{\pm}}$}
   \includegraphics[width=1.0\linewidth]{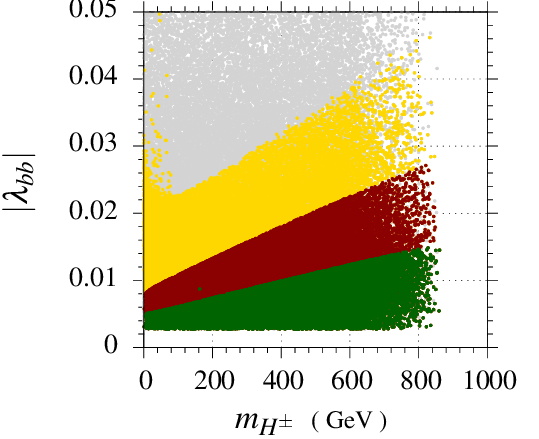}
   \label{fig13:sfig4}
 \end{subfigure}%
 \caption{Effect of constraints on $|\lambda_{bb}|$ vs $m_{H^{\pm}}$ plane. Color coding is same as in
Fig. \ref{fig:combined}}
 \label{fig:13}
 \end{figure}

The constraints on $|\lambda_{bb}|$ with $m_{H^{\pm}}$ from above mentioned $B$-mesons decays are plotted in Fig. \ref{fig:13}. In 
Fig. \ref{fig13:sfig1}, we can observe that from $B_{s} \to \mu^{+} \mu^{-}$ decay, in case wen 
$m_{H^{\pm}}> 85$ GeV, all the range of $|\lambda_{bb}|\simeq 0.05$ is allowed. However, when we brought in the $q^2_{0}$, there is
a linear increase in the allowed range of $|\lambda_{bb}|$ as we increase $m_{H^{\pm}}$. One can see that $|\lambda_{bb}|\simeq 0.028$ is the maximum possible value for $m_{H^{\pm}}\simeq 800$ GeV. Similarly, when we tried to put constraints from $b \to s \gamma$ decay, again the trend of $|\lambda_{bb}|$ is linearly increasing and 
$|\lambda_{bb}|\simeq 0.016$ for $m_{H ^ {\pm}} \simeq 800$ GeV.
Now for Type-II of the 2HDM, in Fig. \ref{fig13:sfig2}, the upper bound on $|\lambda_{bb}|$ is 
$|\lambda_{bb}|\simeq 0.27$. Same is the case for Type-III as can be seen in Fig. \ref{fig13:sfig3}. 
 In case of the Type-IV, in Fig. \ref{fig13:sfig4}, the constraints from $q^2_{0}$ and $b \to s \gamma $ are 
 similar to those in Fig. \ref{fig13:sfig1}. But when constraints come from the  $B_{s} \to \mu^{+}\mu^{-}$ there is also a 
 linear increase in the allowed range of $|\lambda_{bb}|$ for the relative increment in the 
 $m_{H ^{\pm}}> 85$ GeV and the maximum possible value is
 $|\lambda_{bb}|\simeq 0.046$.  
 
\begin{figure}[H]
 \begin{subfigure}{.5\textwidth}
   \centering
 	\caption{Type I $|\lambda_{bb}|$ vs $m_{H}$}
   \includegraphics[width=1.0\linewidth]{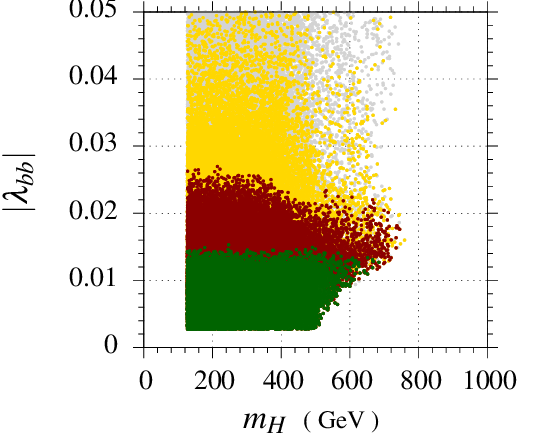}
   \label{fig14:sfig1}
 \end{subfigure}%
 \begin{subfigure}{.5\textwidth}
   \centering
   \caption{Type II $|\lambda_{bb}|$ vs $m_{H}$}   
   \includegraphics[width=1.0\linewidth]{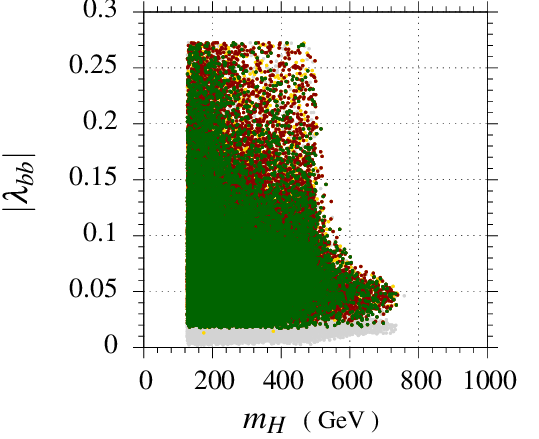}
   \label{fig14:sfig2}
 \end{subfigure}%
 \\
 \begin{subfigure}{.5\textwidth}
   \centering
   \caption{Type III $|\lambda_{bb}|$ vs $m_{H}$}
   \includegraphics[width=1.0\linewidth]{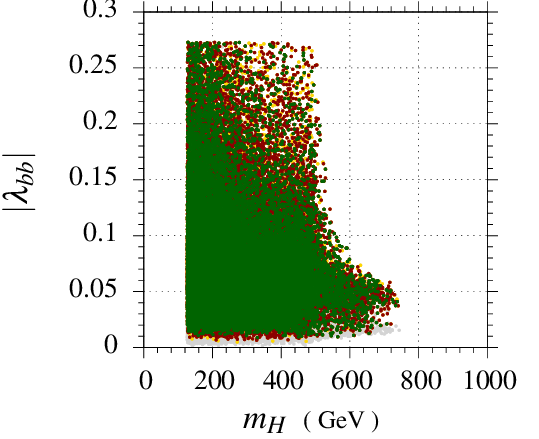}
   \label{fig14:sfig3}
 \end{subfigure}%
 \begin{subfigure}{.5\textwidth}
   \centering
   \caption{Type IV $|\lambda_{bb}|$ vs $m_{H}$}
   \includegraphics[width=1.0\linewidth]{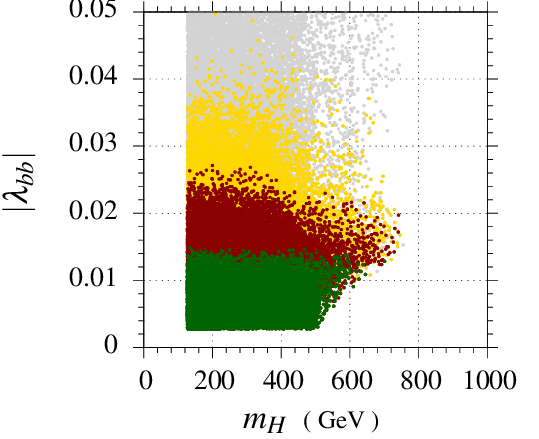}
   \label{fig14:sfig4}
 \end{subfigure}%
 \caption{Effect of constraints on $|\lambda_{bb}|$ vs $m_{H}$ plane. Color coding is same as in
Fig. \ref{fig:combined}}
 \label{fig:14}
 \end{figure}

 We next show our results in $|\lambda_{bb}| - m_{H}$ plane in Fig. \ref{fig:14}. In case of the Type-II and 
 Type-III of the 2HDM, (c.f. Figs. \ref{fig14:sfig2} and \ref{fig14:sfig3}, respectively) the upper bound on $|\lambda_{bb}|$ is
 $|\lambda_{bb}|\simeq 0.27$ for $m_{H} \simeq 550$ GeV. However, for $m_{H}>550$ GeV this allowed range 
 decreases very sharply and for the highest allowed value of $m_{H}$ in our analysis, i.e., $m_{H}\simeq750$ GeV, 
 $|\lambda_{bb}|= 0.05$ is fixed. There is a slightly lower bound on $|\lambda_{bb}|$ for these
 two types as for Type-II $|\lambda_{bb}|\simeq 0.02$ and for Type-III it is 
 $|\lambda_{bb}|\simeq 0.01$.
In case of the Type-I and Type-IV of the 2HDM, trends of $B$-Physics observables are almost similar. 
$B_{s} \to \mu^{+}\mu^{-}$ has very marginal effect on these plots. However, $q^2_{0}$ allows 
$|\lambda_{bb}|\simeq 0.026$ for $m_{H}\leq 400$ GeV and this limit slightly decreases 
$|\lambda_{bb}| \simeq 0.024$ for $400<m_{H}<500$ GeV. For $m_{H}> 500$ GeV the allowed range of 
$|\lambda_{bb}|$ increase linearly up-to $|\lambda_{bb}|\simeq 0.024$.

\begin{figure}[H]
 \begin{subfigure}{.5\textwidth}
   \centering
 	\caption{Type I $|\lambda_{bb}|$ vs $m_{A^{o}}$}
   \includegraphics[width=1.0\linewidth]{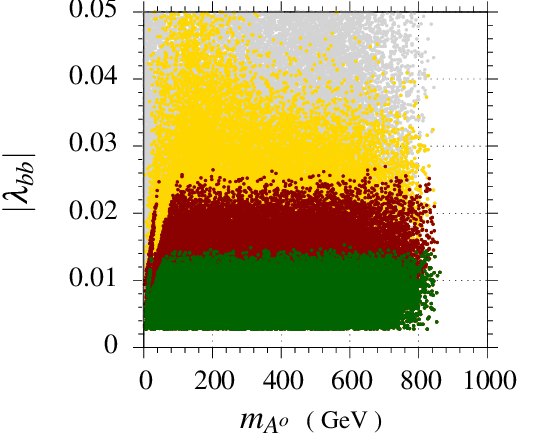}
   \label{fig15:sfig1}
 \end{subfigure}%
 \begin{subfigure}{.5\textwidth}
   \centering
  \caption{Type II $|\lambda_{bb}|$ vs $m_{A^{o}}$}
   \includegraphics[width=1.0\linewidth]{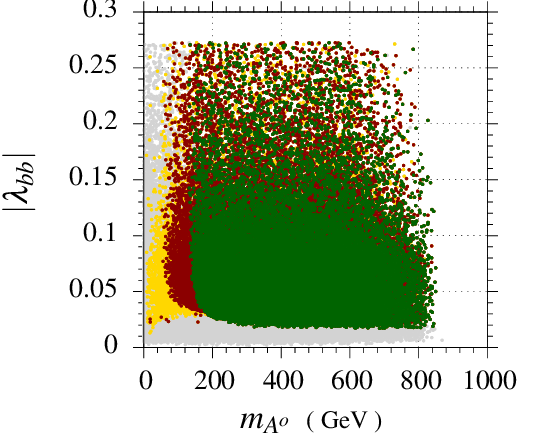}
   \label{fig15:sfig2}
 \end{subfigure}%
 \\
 \begin{subfigure}{.5\textwidth}
   \centering
   \caption{Type III $|\lambda_{bb}|$ vs $m_{A^{o}}$}
   \includegraphics[width=1.0\linewidth]{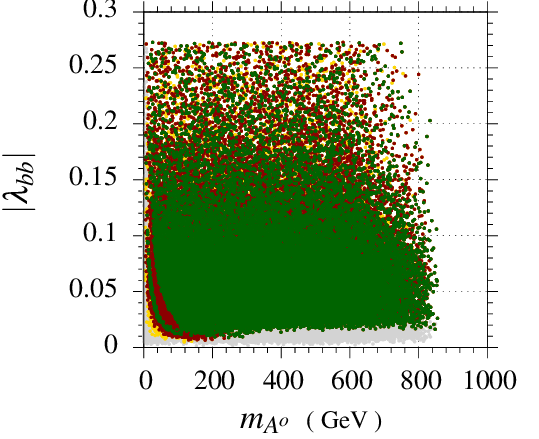}
   \label{fig15:sfig3}
 \end{subfigure}%
 \begin{subfigure}{.5\textwidth}
   \centering
   \caption{Type IV $|\lambda_{bb}|$ vs $m_{A^{o}}$}
   \includegraphics[width=1.0\linewidth]{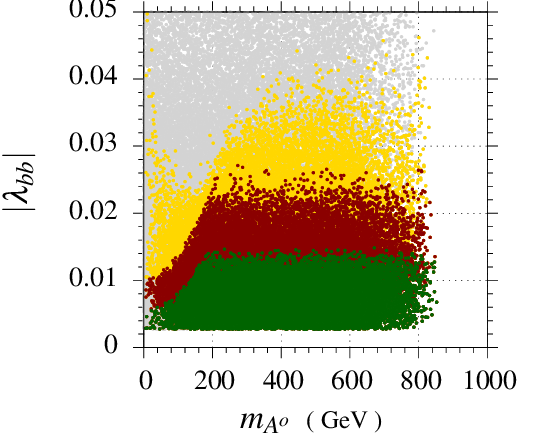}
   \label{fig15:sfig4}
 \end{subfigure}%
 \caption{Effect of constraints on $|\lambda_{bb}|$ vs $m_{A^{o}}$ plane. Color coding is same as in
 Fig. \ref{fig:combined}}
 \label{fig:15}
 \end{figure}

We next present our results in $|\lambda_{bb}| - m_{A^{0}}$ planes for the four types of the 2HDM in 
Fig. \ref{fig:15}. In Fig. \ref{fig15:sfig1}, one can notice that 
$B_{s} \to \mu^{+}\mu^{-}$ has no notable constraint on this plot, whereas $q^2_{0}$ allows 
$|\lambda_{bb}|\leq 0.026$ for the mass range of $m_{A ^ {0}}>80$ GeV. In case of putting 
constrains on $|\lambda_{bb}|$ from  $B \to X_s \gamma$ the allowed range is $|\lambda_{bb}|\leq 0.014$.
Figs. \ref{fig15:sfig2} and \ref{fig15:sfig3} display $|\lambda_{bb}| - m_{A^{o}}$ plane for Type-II and 
Type-III of the 2HDM. The upper limit on $|\lambda_{bb}|$ is $|\lambda_{bb}|\simeq 0.27$ which is ten 
times higher than the allowed limit for $|\lambda_{bb}|$ in Type-I by $q^2_{0}$ and almost nineteen
times higher than allowed limit of $|\lambda_{bb}|$ by $b \to s \gamma$. There is also some 
difference of humps, which corresponds to lower limit on $|\lambda_{bb}|$, between Type-II and 
Type-III planes for $m_{A ^{\circ}}\leq 350$ GeV. In case of the Type-IV, Fig. \ref{fig15:sfig4}, 
constraints from $q^2_{0}$ and $b \to s \gamma$ have similar effects to that of the Type-I in 
Fig. \ref{fig15:sfig1}. For $B_{s} \to \mu^{+} \mu^{-}$, the trends are different for 
$m_{A ^{\circ}}\leq 400$ GeV.   
%
\section{Conclusions}
In this work we scanned the parametric space of  all four types of the 2HDM by imposing $Z_{2}$ symmetry on Yukawa Lagrangian and by incorporating the experimental constraints from the different 
observables of rare $B$-meson decays. In addition to these conditions,
the theoretical constraints from the unitarity of $S$-martix and positivity of the potential of 2HDM are also incorporated. The observables which were taken into account
to constrained the 2HDM parameters are the branching ratios of $B_{s}\to X_{s}\gamma$, $B_{s}\to\mu^{+}\mu^{-}$ and zero crossing of the forward-backward
asymmetry in $B\to K^{\ast}\mu^{+}\mu^{-}$ decay. Among these three observables the most stringent constraints are coming from the branching ratio of 
$B_{s}\to X_{s}\gamma$ that is now studied at two loops theoretically in ref. \cite{Misiak:2015xwa}. The main outcomes of our study can be summarized as follows:
\begin{itemize}
 \item It is observed that for all  types of the 2HDM, the upper limit of charged Higgs boson $m_{H^{\pm}}\sim 840$ GeV. For type-I of the 2HDM in $m_{H^{\pm}}-\tan\beta$ plane,
 it is found from the experimental value of branching ratio of the decay $B_{s}\to\mu^{+}\mu^{-}$ that the lower bound of
 $m_{H^{\pm}}> 80$ GeV and is in agreement with LEP data \cite{Agashe:2014kda} for $\tan\beta\sim 2$. However for type-II of 2HDM, in 
 the same plane, it is found from the experimental value of $B_{s}\to X_{s}\gamma$ that the allowed range of $m_{H^{\pm}}$ 
 is $(460$ \text{GeV}$\leq m_{H^{\pm}}\leq 840$ \text{GeV}$)$ and this bound agrees with the theoretical calculations done by Misiak \textit{et. al.} \cite{Misiak:2015xwa}.
 \item For $CP$ even Higgs boson mass $m_{H}$, the upper limit is 500 GeV and for $CP$ odd Higgs boson mass $m_{A^{0}}$, the upper bound is 840 GeV in all 
 the four types of the 2HDM. In case of $CP$-even Higgs boson mass it is observed that by increasing the value of $\tan\beta$, upper bound on $m_{H}$ decreases.
 \item From our analysis of $\lambda_{tt}-m_{H^{\pm}}$, $\lambda_{tt}-m_{H}$ and $\lambda_{tt}-m_{A^{0}}$ planes it can be observed that there 
 are severe bounds on $\lambda_{tt}$ from the radiative decay of $B$-meson and its value cannot be greater than 1 for Type-I and Type-IV of the 2HDM. However,
 for Type-II of the 2HDM the upper bound on $\lambda_{tt}$ is 1.55 but for Type-III there are no constraints on the upper bounds of $\lambda_{tt}$.
 \item From our results for $|\lambda_{bb}|$ we infer that the upper bound on $|\lambda_{bb}|$ is 0.015 from the branching ratio of the decay
 $B\to X_{s}\gamma$ for Type-I and Type-IV of 2HDM. Furthermore, the upper bound on $|\lambda_{bb}|$ in Type-II and Type-III of the 2HDM is $0.27$ that is an order of magnitude larger than Types I and IV of the 2HDM. 
 \end{itemize} 
  In our analysis we used the current data of LHCb for the branching ratios of radiative and leptonic decays of $B$ meson and zero crossing of forward backward 
  asymmetry for the decay $B\to K^{\ast}\mu^{+}\mu^{-}$ to predict the allowed ranges of 2HDM parameters. We hope that the findings of the present study could be tested in the future data from LHC.
  \section*{Acknowledgement}
  Authors would like to thank Research Centre for Modeling and Simulations (RCMS) at National University of Sciences and Technology (NUST), Islamabad, Pakistan for providing their super-computing facility in calculating the results used in this article.
  MU is supported by National University of Sciences and Technology (NUST), Sector H-12 Islamabad 44000, Pakistan and Higher Education Commission (HEC) of Pakistan under the project no. NRPU-3053.
\medskip
\bibliographystyle{unsrt}
\bibliography{mybib}
\end{document}